# Common Traffic Congestion Features studied in USA, UK, and Germany employing Kerner's Three-Phase Traffic Theory

Hubert Rehborn, Sergey L. Klenov*, Jochen Palmer#

Daimler AG, HPC: 050-G021, D-71059 Sindelfingen, Germany,
e-mail: hubert.rehborn@daimler.com,
Tel. +49-7031-4389-594, Fax: +49-7031-4389-210

*Moscow Institute of Physics and Technology, Department of Physics, 141700 Dolgoprudny, Moscow Region, Russia,
e-mail: sergey_klenov@yandex.ru

#IT-Designers GmbH, Germany,
e-mail: jochen.palmer@it-designers.de

**Corresponding author**: Hubert Rehborn, e-mail: hubert.rehborn@daimler.com

*Abstract* - Based on a study of real traffic data measured on American, UK and German freeways common features of traffic congestion relevant for many transportation engineering applications are revealed by the application of Kerner's three-phase traffic theory. General features of traffic congestion, i.e., features of traffic breakdown and of the further development of congested regions, are shown on freeways in the USA and UK beyond the previously known German examples. A general proof of the theory's statements and its parameters for international freeways is of high relevance for all applications related to traffic congestion.

The application ASDA/FOTO based on Kerner's three-phase traffic theory demonstrates its capability to properly process raw traffic data in different countries and environments.

For the testing of Kerner's "line J", representing the wide moving jam's downstream front, four different methods are studied and compared for each congested traffic situation occurring in the three countries.

## 1 INTRODUCTION

In many countries of the world, traffic on freeways is often heavily congested during many hours of the day. Common traffic congestion features, especially features of traffic breakdown as well as features of propagating "jam" structures, have to be taken into account in diverse applications, e.g., adaptive cruise control systems, traffic safety applications, V2V / V2I (Vehicle-to-Vehicle / Vehicle-to-Infrastructure) technologies for individual vehicles as well as traffic control and management systems for collective traffic management centres.

In recent years the so-called three-phase traffic theory has been proposed by Kerner (2004, 2009b): in addition to free flow traffic phase (F), the lower speed states of congested traffic on freeways have to be distinguished between the two traffic phases: synchronized flow (S) and wide moving jam (J). While the synchronized flow regions remain often fixed at the location of the bottleneck and the wide moving jam propagates through any kind of a bottleneck, both congested traffic phases might have similar vehicle speeds, e.g., measured by local detectors. Hence, only a spatial-temporal investigation of a congested traffic state allows the consistent congested traffic phase classification.

The majority of today's traffic models, control and management approaches as well as its vehicular applications have been criticized by the three-phase traffic theory in Kerner (2004, 2009b) mainly for two reasons: features of traffic breakdown as well as the further development of the related congested region have not been understood properly. In the earlier models with traffic flow instability (e.g., review by Helbing (2001)), traffic breakdown was explained by a phase transition from free flow to wide moving jams ($F \rightarrow J$ transition). In contrast, in real traffic flow traffic breakdown is governed by a phase transition from free flow to synchronized flow ($F \rightarrow S$ transition) (Kerner (1997); with examples in the paper).

A correct and efficient traffic data analysis is a principle task in traffic engineering (e.g., Treiterer (1975); Cremer (1979); Leutzbach (1988); May (1990); Daganzo (1997); Helbing (2001); Highway Capacity Manual (2000); Howe (2006); Kerner and Rehborn (1996a, 1996b); Kerner (2004, 2009a, 2009b)). Kerner's three-phase traffic theory gives a foundation for understanding the phenomena of freeway congestion. Practical applications like the models ASDA (devoted to the automatic tracking of wide moving jams) and FOTO (devoted to the traffic phase classification and the tracking of the synchronized flow phase) (Kerner (1999a, 2004, 2009b); Kerner and Rehborn (1998), Kerner et al. (1998, 1999, 2004)) are based on Kerner's theory.

Especially the wide moving jams have characteristic parameters which should be derived efficiently based on given data sets (e.g., velocity of the jam's downstream front, flow rate out of the wide moving jam in case of downstream free flow). Some methods for the determination of these parameters will be analysed in this paper for all investigated countries using typical and regular examples of traffic congestions.

Kerner's three-phase traffic theory questions many results of earlier traffic flow theories and models reviewed for example in Leutzbach (1988); May (1990); Highway Capacity Manual (2000); Helbing (2001). Therefore, it is not surprising that Kerner's theory has been criticized by many authors, e.g., by Daganzo et al. (1999); by Treiber et al. (2000, 2010) and by Schönhof and Helbing (2007, 2009). Critical responses to these and other criticisms of three-phase traffic theory have already been published in chapter 10 of the book by Kerner (2009b) in which a critical analysis of earlier traffic flow models and theories has been made. In this solely *empirical* paper, we do not discuss this controversial *theoretical* subject and refer readers to the abovementioned chapter 10 of Kerner (2009b).

Kerner's three-phase traffic theory as well as its empirical basis have already been reviewed in both books (Kerner (2004, 2009b)). However, the empirical basis of the theory presented in these books is mainly related to traffic data measured on German freeways. Thus, the task arises to test this theory with traffic data of other countries. Without such a test, practical applications of this theory might be very limited. We believe that in this paper for the first time common features of traffic congestion are revealed based on real measured data on freeways in USA, UK, and Germany. Rather than a comprehensive review of Kerner's theory made already in his books, the main novel aims of this review paper are as follows:

1. The illustration of definitions and results of Kerner's theory with representative empirical examples from three different countries which show commonalities investigated in several years of traffic data observations in Germany, approximately one year of traffic data from UK's M42 freeway and about a month of US data from Oregon.
2. The comparison of different methods for obtaining moving jam characteristics related to Kerner's theory.

However, before we start with novel results, we should recall some definitions and results of Kerner's theory needed for the paper understanding.

## 2 BACKGROUND: BASIC DEFINITIONS AND SOME RESULTS OF KERNER'S THREE-PHASE TRAFFIC THEORY

### 2.1. Definitions [J] and [S] for traffic phases in congested traffic

Freeway traffic can be either free or congested (e.g., Leutzbach (1988); May (1990); Highway Capacity Manual (2000); Helbing (2001)). In empirical observations, congested traffic is often defined as traffic in which averaged vehicle speed is lower than the minimum possible average speed in free flow (see for example Helbing (2001)). Congested traffic is mostly observed at bottlenecks (e.g., Leutzbach (1988); May (1990); Highway Capacity Manual (2000); Helbing (2001)).

Based on extensive traffic data analyses of available stationary measurements spanning several years Kerner discovered that two different traffic phases must be differentiated in congested freeway traffic: "synchronized flow" and "wide moving jam" (Kerner (1997, 1999b, 2004); Kerner and Rehborn (1996a, 1996b, 1997)). Thus, there are three phases in this theory: free flow (F), synchronized flow (S), and wide moving jam (J). Empirical macroscopic spatiotemporal criteria for congested traffic phases as elements of Kerner's three-phase traffic theory are as follows:

The definition of the wide moving jam phase [J]: A wide moving jam is a moving jam that maintains the mean velocity of the downstream jam front, even when the jam propagates through any other traffic state or a freeway bottleneck. This is the characteristic jam feature *J*.

The definition of the synchronized flow traffic phase [S]: In contrast, the downstream front of the synchronized flow phase does not show the characteristic jam feature; in particular, the downstream front of synchronized flow is often fixed at the bottleneck.

It must be noted that in Kerner's theory *neither* the observation of speed synchronization in congested traffic *nor* other relationships and features of congested traffic measured at specific freeway locations (e.g., in the flow-density plane) are some criteria for the phase differentiation in congested traffic. The clear differentiation between the synchronized flow and wide moving jam phases can be made on the above objective criteria [J] and [S] *only*.

The empirical phase definitions [J] and [S] are illustrated in Figure 1(a) in which real measured traffic data with two different congested patterns are shown: The first pattern propagates through a bottleneck while maintaining the mean velocity of its downstream front. In accordance with the phase definition [J], the pattern is an example of the wide moving jam phase. In contrast, the downstream front of another congested pattern is fixed at the bottleneck. In accordance with the phase definition [S], the pattern is an example of the synchronized flow phase.

### 2.2. The fundamental hypothesis of three-phase traffic theory

The fundamental hypothesis of Kerner's three-phase traffic theory is as follows: Steady states of synchronized flow cover a two-dimensional (2D) region in the flow-density plane (so-called "Kerner's 2D-states") (Figure 1(b)). A steady state of synchronized flow is hypothetical homogeneous in space and time synchronized flow in which all vehicles move at the same speed and the same space (and time) headways. This means that in contrast to the hypothesis of earlier traffic flow theories there is no fundamental diagram for steady states of traffic flow in this theory.

*2.3. Traffic breakdown*

As it is well-known, congested traffic occurs in an initial free flow due to traffic breakdown, i.e., a sharp decrease in vehicle speed. This breakdown occurs mostly at a bottleneck (e.g., Leutzbach (1988); May (1990); Highway Capacity Manual (2000); Helbing (2001)). In Kerner's three-phase traffic theory traffic breakdown at a bottleneck is associated with a phase transition from the free flow phase [F] to the synchronized flow phase [S] (called $F \rightarrow S$ transition).

An empirical example of traffic breakdown is shown in Figure 1(a): In this example, a wide moving jam propagating through a bottleneck causes traffic breakdown with the emergence of synchronized flow. Such traffic breakdown, i.e., $F \rightarrow S$ transition can be considered "induced" traffic breakdown because the breakdown at the bottleneck is induced by the wide moving jam when the jam has reached the bottleneck.

There can be also a "spontaneous" traffic breakdown at the same bottleneck caused by a local disturbance within initial free flow phase at the bottleneck, i.e., when before the breakdown has occurred there have been initially free flows at the bottleneck as well as both upstream and downstream of the bottleneck.

*2.4. Kerner's line J and phase transition from synchronized flow to wide moving jam*

The line J (so-called "Kerner's line *J*") represents the velocity of the downstream front of a wide moving jam in the flow-density plane (line *J* in Figure 1(b)).

In empirical measured wide moving jams, typically the jam velocity related to the line *J* with approximately $-15$ km/h has been found (angle $v_{gr}$ in Figure 1(b)). The line *J* that represents the downstream front of a wide moving jam starts at the jam density $\rho_{max}$ from the standstill of vehicles (in measurements the jam density is about 140 vehicle/km for passenger cars).

Kerner's line *J* has also another important feature: The line *J* divides states of synchronized flow (2D-region labelled by "S" in Figure 1(b)) into two different classes:

(i) Synchronized flow states *on* and *above* the line *J* are metastable regarding the phase transition from synchronized flow to wide moving jam ($S \rightarrow J$ transitions), i.e., such phase transitions might occur due to a traffic disturbance.

(ii) In synchronized flow states *below* the line *J* no wide moving jam emergence is possible, because they will dissolve over time.

Additionally, if free flow is formed in the outflow of a wide moving jam, the flow rate $q_{out}$ in this jam outflow is much lower than the maximum flow rate in free flow $q_{max}^{(free)}$. In empirical observations of 1-min average data a relation of $q_{max}^{(free)} / q_{out} \approx 1.5$ has been found in Kerner and Rehborn (1996b). If there is no free flow in the outflow of the wide moving jam, the line *J* ends within the traffic phase [S].

*2.5. Critical disturbances of traffic flow for phase transitions*

Because there are three mentioned traffic phases [F] (free flow), [S] (synchronized flow) and [J] (wide moving jam) in Kerner's theory, there can occur any phase transition between these three traffic phases.

Each of these phase transitions occurs when a critical local disturbance in an initial traffic phase appears whose amplitude exceeds some critical value. Examples for practical reasons of such critical disturbances in free and synchronized flows could be overtaking manoeuvres, fluctuations in flow rates upstream, a sudden vehicle braking, etc.. Critical amplitudes of these critical disturbances required for phase transitions and the dependencies of the critical disturbance amplitudes on traffic variables (e.g., on the vehicle density) are *very different* for different phase transitions. In particular, amplitudes of critical disturbances leading either to traffic breakdown at a bottleneck, i.e., the phase transition from free flow to synchronized flow ($F \rightarrow S$ transition), or to a phase transition from free flow to wide moving jam ($F \rightarrow J$ transition), or else to a phase transition from synchronized flow to wide moving jam ($S \rightarrow J$ transition) are qualitatively illustrated in Figure 1(c).

In free flow, at the same flow rate density the amplitude for the $F \rightarrow S$ transition is much lower than for the $F \rightarrow J$ transition (see curves $F_J$ and $F_S$ in Figure 1(c)). This explains why traffic breakdown at a bottleneck in all real measured traffic data is the $F \rightarrow S$ transition rather than the $F \rightarrow J$ transition.

Similarly, the amplitude of a critical perturbation in synchronized flow of a speed $v_{syn}$ has a maximum exactly on the line *J* and decreases to the upper boundary of the two-dimensional synchronized flow region (curve $S_J$ in Figure 1(c)).

It must be noted that traffic breakdown, i.e., a $F \rightarrow S$ transition occurs usually at a bottleneck. After synchronized flow has emerged due to the breakdown, a wide moving jam(s) can occur in synchronized flow ($S \rightarrow J$ transition). Thus, wide moving jams emerge later than traffic breakdown has occurred at the bottleneck and at another road location that the breakdown location.

*2.6. On-line applications of ASDA/FOTO in real installations*

One of the engineering applications of the three-phase traffic theory are the models ASDA and FOTO proposed by Kerner and further developed by Rehborn, Aleksic, Haug and Palmer for recognition and tracking of spatiotemporal congested traffic patterns as described in Kerner (1999a, 2004, 2009a, 2009b); Kerner and Rehborn (1998); Kerner et al. (1998, 1999, 2004).

Within the FOTO model, at first the traffic phase identification is performed. Based on, e.g., loop detectors, the local measured average velocities and flow rates are classified into traffic phases: If velocity is low and flow rate is high, the traffic phase is "synchronized flow". If both velocity and flow rate are low, the phase is "wide moving jam". In all other cases, the phase is "free flow". Secondly, FOTO performs the recognition of the locations of the upstream and downstream fronts $x_{up}^{(syn)}$, $x_{down}^{(syn)}$ of synchronized flow. It must be noted that the traffic phase identification in ASDA/FOTO is *not* based on the phase definitions [J] and [S]. The reason for this is as follows: The application of the definitions [S] and [J] could be possible only after the spatiotemporal structure of a congested pattern has already been known. In contrast, ASDA/FOTO should make online traffic phase identification at each time instant, i.e., also before this spatiotemporal pattern structure is known. In other words, the traffic phase identification used in ASDA/FOTO is only an *approximate* one.

While the downstream front $x_{down}^{(syn)}$ of synchronized flow is fixed at a bottleneck, the position of the upstream front $x_{up}^{(syn)}$ with respect to the bottleneck is calculated by a "cumulative flow" approach with the inflowing vehicles $q_0(t)$ [vehicles/h] and the vehicles escaping downstream from the synchronized flow $q_n(t)$ [vehicles/h] in Kerner et al. (2004):

$$x_{up}^{(syn)}(t) = \mu \frac{1}{n} \int_{t_{syn}}^{t} (q_n(t) - q_0(t))dt \quad t \geq t_{syn} \tag{1}$$

with $20 < \mu < 40$ [m/vehicles] as parameter, n as the number of freeway lanes and $t_{syn}$ as a time at which the synchronized flow is first registered.

Then, the ASDA model recognizes the upstream and downstream fronts $x_{up}^{(jam)}$, $x_{down}^{(jam)}$ of wide moving jams. The positions of these jam fronts over time are calculated by:

$$x_{up}^{(jam)}(t) = \int_{t_0}^{t} v_{gl}(t)dt \approx -\int_{t_0}^{t} \frac{q_0(t) - q_{min}}{\rho_{max} - (q_0(t)/v_o(t))}dt, \quad t \geq t_0 \tag{2}$$

$$x_{down}^{(jam)}(t) = \int_{t_1}^{t} v_{gr}(t)dt \approx -\int_{t_1}^{t} \frac{q_n(t) - q_{min}}{\rho_{max} - (q_n(t)/v_n(t))}dt, \quad t \geq t_1 \tag{3}$$

with $q_0(t), q_n t)$ as the upstream (downstream) flow rates, $v_0(t), v_n t)$ as the upstream (downstream) vehicle speeds, $q_{min}$ as the flow rate in the wide moving jam (normally set to zero) and the parameter $\rho_{max}$ as the density inside the wide moving jam, $t_0, t_1$ are time moments at which the upstream and the downstream fronts are first registered at the downstream position, respectively. The densities $\rho_{max}$ and $\rho_{min} = q_n(t)/v_n(t)$ in case when free flow is formed downstream of the jam are the two distinct characteristic densities that determine the velocity of the downstream front of a wide moving jam.

The formulas (2), (3) are associated with the classic Stokes formula for the shockwave velocity $v_s$ as follows:

$$v_s = \frac{q_2 - q_1}{\rho_2 - \rho_1} \tag{4}$$

with $q_1, \rho_1$ as the flow rate and density downstream of the shockwave and $q_2, \rho_2$ as the flow rate and density upstream of the shockwave. It must be stressed that this Stokes formula Eq. (4) and the related ASDA formulas Eq. (2) and (3) are fundamentally different from the well-known formula for shockwaves in the classic Lighthill-Whitham-Richards (LWR)-theory by Whitham (1974):

$$v_s^{(LWR)} = \frac{Q(\rho_2) - Q(\rho_1)}{\rho_2 - \rho_1} \tag{5}$$

where $Q(\rho_2)$ and $Q(\rho_1)$ are flow rates on the fundamental diagram that gives a single correspondence between the density and the flow rate. In contrast to Eq. (5) in ASDA formulas Eqs. (2) and (3) there is no given correspondence between the density $\rho_2$ and the flow rate $q_2$ as well as between the density $\rho_1$ and the flow rate $q_1$. ASDA solves the problem to find $q_2 = q_0$ in Eq. (2) through the use of measured data from an upstream detector. In turn, $q_1 = q_n$ in Eq. (3) is found through the use of measured data at the downstream detector.

For each time interval (e.g., 1 minute) based on measured data, ASDA formulas Eq. (2) and (3) find the density as a ratio of the flow rate and the speed. As shown in empirical observations in synchronized flow there is no single relationship between flow rate and density. Instead there are an infinite number of such relations covering a two-dimensional region in the flow-density plane. In other words, in contrast to the LWR-theory, there is no fundamental diagram of congested traffic that gives a single flow rate for a given density or vice versa a single density for a given flow rate.

The main feature of the ASDA and FOTO models is that the models identify firstly two different traffic phases in congested traffic (synchronized flow and wide moving jam), and then synchronized flow is tracked with the "cumulative flow" approach, while in contrast wide moving jams are tracked with the Stokes shockwave formula.

The front locations $x_{up}^{(syn)}$, $x_{down}^{(syn)}$ and $x_{up}^{(jam)}$, $x_{down}^{(jam)}$ define the spatial size and location of the related "synchronized flow" and "wide moving jam" objects, respectively. Finally, the ASDA and FOTO models track these object fronts $x_{up}^{(syn)}(t), x_{dwon}^{(syn)}(t), x_{up}^{(jam)}(t), x_{dwon}^{(jam)}(t)$ in time and space (see Figure 2). Note, that through the use of the ASDA and FOTO models the tracking of congested traffic objects is also carried out between detectors, i.e., when the object fronts cannot be measured at all. Additionally, the ASDA and FOTO models work without any validation of model parameters in different environmental and traffic conditions. Model applications are not limited to stationary detector measurements which could measure the necessary flow rates and vehicle speed directly; the use of more advanced measurement technologies like floating car data (vehicles acting as moving traffic sensors) or phone probes (phones acting as moving traffic sensors) will also be possible.

## 3  FREEWAY INFRASTRUCTURES AND MEASUREMENTS

### 3.1 Freeway M42

The UK Highways Agency has implemented an Active Traffic Management (ATM) system as a pilot scheme over the 17km stretch between junctions "3a" and "7" of the M42 (see Figure 3) close to Birmingham with many loop detectors. The freeway M42 has three lanes and about 130 detectors between intersections "3a" and "5" (Figure 3). Each detector measures flow rates and average speeds per lane in one minute intervals. A traffic control system is capable of showing variable message signs to the drivers. The dynamic traffic data is available at the traffic control centre near Birmingham.

### 3.2 Interstate I5

In the US state of Oregon a roadside infrastructure around Portland offers data from some hundreds of loop detectors in a web portal since more than five years (see: PORTAL website http://portal.its.pdx.edu/index.php). The Interstate I5 near Portland has two to three lanes and 8 freeway detector stations named D16 to D23 on the approx. 10km section (see Figure 4). Each detector measures flow rates and occupancy rates in 20sec intervals. All dynamic traffic data is available via Portland PORTAL website.

### 3.3 Freeway A5

In Germany, the freeway data from the A5 with a variable message sign control system has been extensively investigated in recent years (e.g., Kerner (1997, 1999b, 2004, 2009a, 2009b); Kerner and Rehborn (1996a, 1996b, 1997); Kerner et al. (2004); Rehborn and Klenov (2009)). The freeway A5 near Frankfurt has generally three lanes and about 30 detectors on the 30 km stretch between "Westkreuz Frankfurt" and north of "Friedberg" (Figure 5). Each detector measures flow rates and average speeds per lane in one minute intervals. All dynamic traffic data is available at the traffic control centre near Frankfurt.

## 4 TRAFFIC BREAKDOWN

In accordance with Kerner's three-phase traffic theory, we have found that traffic breakdown is a $F \rightarrow S$ transition at a bottleneck. This phase transition can clearly be seen in measured traffic data on UK freeway M42 and US Portland's I5-North (labelled by arrow $F \rightarrow S$ in Figure 6(b) and Figure 7(b)). All measured features of these phase transitions are qualitatively the same as those found in German traffic data in Kerner (2004).

At the bottleneck named "B" in Figures 6(a) and 7(a) a $F \rightarrow S$ transition, i.e., traffic breakdown has occurred. Later, a $S \rightarrow J$ transition might occur in synchronized flow leading to a wide moving jam propagating further upstream. Each of the Figures 6(b) and 7(b) give average velocities (in km/h) on the left and flow rates (in vehicles/h) on the right for the detectors of Figures 6(a) and 7(a), respectively.

Both empirical examples illustrate that traffic breakdown on freeways is a $F \rightarrow S$ transition as stated in the three-phase traffic theory.

## 5 MOVING JAM EMERGENCE

After synchronized flow has occurred as a result of the $F \rightarrow S$ transition, the upstream front of the synchronized flow propagates upstream, whereas the downstream front is fixed at the bottleneck. During the upstream synchronized flow propagation, a strong compression of this synchronized flow is observed (see detector measurements in 100 m distances in Figure 8 for M42). This compression of the synchronized flow means that the speed decreases strongly whereas the flow rate remains almost the same. Hence, the density of synchronized flow increases strongly upstream of the bottleneck. In this dense synchronized flow a narrow moving jam emerges. The jam propagates upstream growing in its amplitude; as a result of the jam growth a wide moving jam emerges; this is called $S \rightarrow J$ transition. The compression of synchronized flow with the subsequent moving jam emergence is called the pinch effect in synchronized flow (see Kerner (2004)). As a result of the pinch effect, a congested pattern with a wide moving jam appears upstream of the bottleneck.

Therefore, these congested traffic situations at freeways M42 and I5-North support the following statements of three-phase traffic theory:
1. Traffic breakdown is a $F \rightarrow S$ transition
2. All wide moving jams occur in synchronized flow, i.e., due to a $F \rightarrow S \rightarrow J$ phase transition
3. $F \rightarrow S$ transitions occur at bottlenecks; $S \rightarrow J$ transitions might occur later and further upstream.

Thus just like measured data on German freeways, in measured data on UK and US freeways, wide moving jams emerge due to a cascade of $F \rightarrow S \rightarrow J$ phase transitions.

The pinch effect can be observed on freeway M42 upstream of detector 6429 (Figure 8): detectors at 100 m distances allow to measure this further compression of synchronized flow with the emergence of narrow moving jams (see detectors 6427, 6426, 6425, 6424 at which the pinch effect is labelled by "pinch effect" in Figure 8). From 500 m upstream of detectors 6429 a wide moving jam begins to emerge inside this pinch region of synchronized flow.

## 6 APPROACHES FOR DETERMINATION OF KERNER'S LINE *J*

The ASDA/FOTO online application provides a visualization of the results as a spatiotemporal diagram. Such diagrams showing the freeway locations (y-axis) over time (x-axis) present the two congested traffic phases in different colours, "light grey" for the traffic phase S and "dark grey" for the traffic phase [J]. Horizontal lines mark the locations of detectors, vertical lines the elapsed time in intervals of 15 minutes. The temporal resolution is 1 minute in accordance with the time intervals of the detector measurements.

In the following section the wide moving jam downstream front velocity will be calculated for diverse traffic situations using different methods. Possible approaches for a systematic and automatic determination of the velocity of the wide moving jam's downstream front (angle of line *J* in flow-density plane) are:

(i) <u>Graphical method</u> chooses the spatiotemporal diagram of ASDA/FOTO and estimate the slope of the moving of the downstream jam front in space-time plane, directly determining the downstream front velocity.

(ii) <u>Detector based method</u> chooses at least two detectors at larger distances and calculate the velocity of the front by using detector distance and registration time of the related downstream jam front at the detectors.

(iii) <u>Correlation method</u>: Speed correlations at different locations are used for the estimation of the jam front velocity.

(iv) <u>Flow-density method</u> calculates the parameters of Kerner's line *J*, if free flow is measured after the downstream front, i.e., in the jam outflow. Estimate the wide moving jam's density and calculate the slope of the line *J* in the flow-density plane.

In this section, traffic congestions from the freeways M42, Interstate I5, and A5 are investigated with these methods for diverse congested traffic situations.

*6.1 Graphical Method*

The following space-time diagrams show the propagating structures of wide moving jams. Graphically the slope of the wide moving jams gives an estimation of the velocity of the downstream fronts.

Figure 9 shows several wide moving jams propagating upstream about more than 15 km on the M42 on 11$^{th}$ January, 2008. The black line gives a graphical estimation of the velocity of the wide moving jam's downstream fronts $v_{gr} = -14.5 \, km/h$. Two detectors "6389" and "6310" are labelled as well as the 14:00-15:00h time window.

Figure 10 illustrates a wide moving jam on Interstate I5-North propagating for approximately 4km on 20$^{th}$ February, 2009. The graphical estimation of the velocity of the wide moving jam's downstream front gives $v_{gr} = -15 \, km/h$.

Figure 11 shows several wide moving jams propagating upstream about 25 km on the A5 on 8$^{th}$ August, 2008. All the three lines give a graphically estimated average value of $v_{gr} = -16 \, km/h$.

Figure 12 shows one unique wide moving jam propagating upstream about more than 25km at the A5 on 16$^{th}$ October, 2008. The black line gives here a graphical estimation of the velocity of the wide moving jam's downstream front of $v_{gr} = -17\,km/h$.

The related measured traffic data of averaged speed and flow rate per minute for these four traffic situations are illustrated in Figures 13-16, respectively.

*6.2 Detector based method*

In Kerner and Rehborn (1996b) the velocity of the wide moving jam's downstream front as a characteristic parameter and the slope of Kerner's line *J* have been calculated by the increasing values of speed and flow measurements at this downstream jam front. The increase registered at different detectors gives combined with the detector's distance the downstream jam front velocity.

An approach used since Kerner and Rehborn (1996b) is as follows: at two detector locations, $D_1$ and $D_2$, the related detector distance is divided by the difference of registration times of the downstream front, i.e.:

$$v_{gr} = \frac{Pos(D_1) - Pos(D_2)}{T_{D1}(v_{D1}(t) > 30km/h) - T_{D2}(v_{D2}(t) > 30km/h)} \tag{6}$$

with $Pos(D_1)$, $Pos(D_2)$ as locations of the related detectors in kilometers and $T_{D1}$, $T_{D2}$ as the specific times in minutes when the measured speeds are higher than 30 km/h after a wide moving jam.

Measured traffic data at selected detectors is illustrated in the following figures: wide moving jams are registered at the detectors by sharp decreases in speed (v in km/h) and flow rate (q in vehicles/h). In each of the Figures 13-16 the time moments of the registration of the wide moving jam's downstream fronts are marked by vertical dotted lines.

Table 1 gives the calculated values for all situations according to formula (6).

*6.3 Correlation method*

Mathematically, the speed measurements of $v_{D1}(t)$ can be correlated with $v_{D2}(t)$. This correlation function has been proposed in Munoz and Daganzo (2002) to distinguish a "wave velocity" between detector stations. The cross-correlation function with time intervals of speed and flow measurements, between, e.g., D11 and D20, on the A5 freeway is as follows:

$$Correl(q)_k^{(D11,D20)} = \frac{\frac{1}{n}\sum_{i=1}^{n}(q_i^{D11} - \overline{q}^{D11}) * (q_{i+k}^{D20} - \overline{q}^{D20})}{\sqrt{\frac{1}{n}\sum_{i=1}^{n}(q_i^{D11} - \overline{q}^{D11})^2} * \sqrt{\frac{1}{n}\sum_{i=1}^{n}(q_{i+k}^{D20} - \overline{q}_k^{D20})^2}} \tag{7}$$

with $\overline{q}^{D11} = \frac{1}{n}\sum_{i=1}^{n}q_i^{D11}$, $\overline{q}_k^{D20} = \frac{1}{n}\sum_{i=1}^{n}q_{i+k}^{D20}$ as average flow rates at the related detectors D11 and D20, n as number of time intervals (here: $n = 120$), i as time interval index, and k as time lag (in minutes). A similar cross-correlation function can be given for speed measurements. The graphical representation of the correlation function over wave velocity $u = 60(Pos(D20) - Pos(D11))/k$ (with (1) 60 in the formula as required to get u in km/h, $Pos(D20)$ and $Pos(D11)$ in km, and (2) k as positive and negative) has been chosen here similarly to Figure 9 in Munoz et al. (2002) as correlation over wave velocities in km/h: the maximum of the correlation value gives the wave velocity in the chosen time interval of the input signal.

The cross-correlation function with time intervals of flow measurements between D11 and D20 shows a maximum at about the aforementioned $-16$ km/h. Hence, in this example with one unique wide moving jam the method proposed in, e.g., Coifman and Wang (2005); Munoz and Daganzo (2002); Zielke et al. (2008) gives the same result as formula (6) as used in Kerner and Rehborn (1996b).

The cross-correlation approach becomes more critical for traffic situations with more than one single wide moving jam, i.e., more complex "signal" measurements of the flow rates. As an example, a similar cross-correlation function for the A5 on August, $8^{th}$ with the three wide moving jams following one another is shown in Figure 17: the cross-correlation function does not have a unique maximum at any velocity, but several other wave velocities have higher positive correlation values. Therefore, the success of this correlation method depends strongly on the choice of the input traffic data.

It has to be noted that even in cases when this correlation methods as in, e.g., Coifman and Wang (2005); Munoz and Daganzo (2002); Zielke et al. (2008) gives a unique value for the propagation jam velocity, this velocity is an averaged velocity of both upstream and downstream fronts of all moving jams in the studied data set. In contrast, the detector based method Kerner and Rehborn (1996b) gives distinct mean velocities separately of upstream and downstream fronts of each moving jam. Therefore, the method by Kerner and Rehborn (1996b) reveals more detail information in traffic data analysis of jam propagation velocities than that of Coifman and Wang (2005); Munoz and Daganzo (2002); Zielke et al. (2008).

*6.4 Flow-density method*

For the calculation of the downstream front's velocity of the wide moving jam with flow-density method, an estimation of the vehicle density inside the wide moving jam is necessary. This can be done with the use of empirical measurements of average percentages of long vehicles $A_{LVeh}$. Using a value of $\rho_{max}$ with an estimated passenger car length of $L_{PCars} = 7.5$ m and $L_{LVeh} = 15$ m for long vehicles this results in:

$$\rho_{max} = \frac{1000}{A_{PCars} * L_{PCars} + A_{LVeh} * L_{LVeh}} \tag{8}$$

The angle $v_{gr}$ for the velocity of the downstream front can then be calculated based on the two measured points at $(q_{out}, \rho_{min})$ and $(q_{min}, \rho_{max})$ as:

$$v_{gr} = -\frac{q_{out} - q_{min}}{\rho_{max} - \rho_{min}}. \tag{9}$$

The flow rate inside the wide moving jam $q_{min}$ can be set to zero. In order to find the outflow from the wide moving jam $q_{out}$ the infrastructure of each freeway has to be checked carefully: on the M42 (approx. 2km upstream of detector 6310) and the A5-North (approx. 6km upstream of detector D20) one can see that the next on-ramp upstream of the related detectors is some distance away: therefore, when free flow is forming after the wide moving jams no new vehicles can have squeezed onto the freeways. Hence the flow rate for the next 6-8 minutes is really the outflow $q_{out}$ from the wide moving jam. Similarly, one can calculate

$\rho_{\min}$ from the measured values based on the averaged speeds after the wide moving jam.

For the US-Interstate I5-North with its many on- and off-ramps in Portland, the measured detector values can be influenced by vehicles squeezing onto the freeway: even if we only take the left lane measurements, it is almost impossible to determine the averaged $q_{out}$ over some time intervals precisely. In addition, in the example there is no free flow formed in the outflow of the wide moving jam: therefore, we can only calculate one averaged point on the line J in synchronized flow (see Figure 19 (b)).

In each of the flow-density planes in Figure 19 the flow rates of the free flow for the related complete day are shown as black points (detectors UK: "6310" (rightmost lane), USA: "18" (left lane) and Germany: "D20" (averaged on all three lanes)).

Additionally, we can check the relation $q_{\max}^{(free)} / q_{out} \approx 1.5$ found in Kerner and Rehborn (1996b) for 1 minute data based on these measurements for all three situations. The examples prove the large gap between the maximum flow rates in free flow at the same traffic parameters (e.g., weather, percentage of long vehicles, etc.) to the average flow rate downstream of a wide moving jam if free flow is formed in the jam outflow. The results summarized in table 2 support the previous results found in Kerner and Rehborn (1996b); Kerner (2004).

*6.5 Comparison*

The differences in the downstream front velocities for different wide moving jams measured on different freeways and countries are caused by different traffic parameters: even if all days would be normal weekday afternoons, the percentage of long vehicles would differ. Other traffic parameters like fog, rain, bad visibility conditions, lane width, etc., have their influence on the quantitative values of the downstream jam front and have not been taken into account here. Qualitatively, it has been shown that the wide moving jam's downstream front velocity can be derived by systematic methods in different situations.

It has to be concluded that the velocity of a wide moving jam's downstream front can be measured with all of these different approaches. The graphical diagram of ASDA/FOTO application gives a first good estimation of $v_{gr}$, which can be proved by analyzing the measured raw traffic data. With local detectors the parameter $\rho_{\max}$ cannot be measured directly: therefore, the length of passenger cars and long vehicles leads to an only average estimation of the density inside the jam. Calculating the flow rates and speeds downstream of the wide moving jam to draw the line *J* in the flow-density plane has to take into account that no new vehicles have squeezed onto the freeway in addition to those drivers who have left the wide moving jam.

The results and application of these four approaches can be concluded in table 3 which mentions the advantages and disadvantages of each method.

Interpreting the result, for the calculation of the wide moving jam's downstream velocity, a combination of the graphical ASDA/FOTO method, which gives a rough estimation of the value, in combination with the detector method based on the raw measurements, correctly performed for different pairs of detectors, are the most efficient approaches.

It should be mentioned that the recognition of the wide moving jam's downstream fronts can also be performed based on moving vehicles (FCD: floating car or floating phone data) and not only based on stationary detectors (see Kerner (2009b); Rehborn and Klenov (2009) for further detail): then the features and peculiarities of moving detectors have to be taken into account.

## 7 CONCLUSIONS

1. There are at least the following *common* spatiotemporal traffic features observed in UK, USA, and Germany:
   - Traffic breakdown at different bottlenecks on freeways in UK, USA, and Germany is always a $F \rightarrow S$ transition: As the result of traffic breakdown firstly the synchronized flow phase of congested traffic emerges at a bottleneck.
   - Later a wide moving jam(s) can spontaneously emerge within this synchronized flow ($S \rightarrow J$ transition). This means that wide moving jams emerge spontaneously as a result of a sequence of $F \rightarrow S \rightarrow J$ transitions. In this sequence, the first phase transition ($F \rightarrow S$) has occurred at the bottleneck. In contrast, the second one ($S \rightarrow J$) occurs later in the emergent synchronized flow and usually upstream of the road location at which the $F \rightarrow S$ transition has occurred.
2. The empirical traffic phase definitions [J] and [S] of Kerner's three-phase traffic theory are *common* spatiotemporal traffic features observed in UK, USA and Germany.
3. ASDA and FOTO models are able to process raw traffic data from UK, USA and Germany with comparable quality regarding the interpretation of the structures of spatiotemporal congested traffic patterns.
4. All congested traffic situations show wide moving jam parameters which have been observed on many other freeways and within several congested situations: the self-organizing process leading to the downstream jam front velocity show an average value of the velocity of $-15$ km/h $+5/-3$ km/h. The variance is caused by different traffic parameters like the percentage of long vehicles in traffic flow, weather, lane widths, etc.. The investigated traffic situations of the three countries are representatives for many more examples.
5. Different approaches for the determination of the wide moving jam's downstream front velocity have been investigated.
6. A detector method is the most efficient and direct approach for the determination of the downstream jam front from measured traffic data. The detector method by Kerner and Rehborn (1996b) gives a detail wide moving jam analysis for its front velocities separately for upstream and downstream fronts and is, therefore, more advantageous than the methods based on speed and flow correlation functions as in used in, e.g., Coifman and Wang (2005); Munoz and Daganzo (2002); Zielke et al. (2008).
7. Flow-density method is valid for the investigation of other parameters of wide moving jams, but the traffic data has to be analyzed more carefully taking possible inflows and outflows at ramps into account.
8. The operation of the models ASDA and FOTO on the freeway M42 in UK and Interstate I5 in USA confirms the results previously obtained in Hessen.

## 8 ACKNOWLEDGEMENTS

We would like to thank B.S. Kerner for many fruitful discussions. In addition, we like to thank W. Balke from Hessen Traffic Control Center for the freeway A5 and E. Wilson from Bristol University for UK's M42 traffic data support. Especially we thank Portland Oregon State's former Prof. R. L. Bertini who has given us access to the raw traffic data of the Portland region.

**Tables with captions:**

Table 1: Wide moving jam's downstream front propagating velocities $v_{gr}$ for diverse traffic situations

| Traffic situation | Detector distance | Jam's downstream front propagation time | $v_{gr}$ |
|---|---|---|---|
| UK: M42 11[th], Jan., 2008 | 7.1 km | 29 min | $-14.5$ km/h |
| US: I5-North 20[th], Feb., 2009 | 4.2 km | 16 min | $-16$ km/h |
| Germany: A5 8[th], Aug., 2008 | 13.8 km | 52 min | $-16$ km/h |
| Germany: A5 16[th], Oct., 2008 | 9 km | 33 min | $-16.5$ km/h |

Table 2: Results of flow-density method

| | UK: M42 | US: I5-North | Germany: A5 |
|---|---|---|---|
| Date | 11$^{th}$, Jan., 2008 | 20$^{th}$, Feb., 2009 | 16$^{th}$, Oct., 2008 |
| $q_{out}$ | 1450 veh/h | 1500 veh/h | 1530 veh/h |
| $v_{gr}$ | $-14.5$ km/h | $-15$ km/h | $-16$ km/h |
| $A_{LVeh}$ | 10 % | unknown | 15 % |
| $\rho_{min}$ | 21 veh/km | 20 veh/km | 20 veh/km |
| | | | |
| $\rho_{max}$ | 121 veh/km | 120 veh/km | 116 veh/km |
| $q_{max}^{(free)}$ | 2340 veh/h | 2160 veh/h | 2060 veh/h |
| $\dfrac{q_{max}^{(free)}}{q_{out}}$ | 1.6 | 1.4 | 1.4 |

Table 3: Approaches for determination of the velocity of the wide moving jam's downstream front as well as other flow characteristic related to the jam

| Approach | Advantages | Disadvantages |
| --- | --- | --- |
| **Graphical method** | Fastest method (manually) | Peculiarities and smoothing of ASDA/FOTO inherently given in the diagram. Not a systematic method. |
| **Detector method** | Exact for two correctly chosen detectors in the preciseness of the measurement interval of one minute. The method permits the determination of the velocities of the upstream and downstream jam fronts separately each other. | Depending on detector choice: two with larger distances necessary. Finding exact registration time of the moving jam's front at the related detector. |
| **Flow-density method** | Calculation of further traffic variables related to wide moving jams like $(q_{out}, \rho_{min})$ and $q_{max}^{(free)} / q_{out}$ relation | Problem of <br> - measuring maximum density inside the wide moving jam $\rho_{max}$ <br> - measuring the outflow of the wide moving jam when not measurements of vehicles merging into the freeway through on-ramps and/or leaving the freeway to off-ramps are available |
| **Correlation method** | Quick method for the automatic estimation of some velocity of the jam propagation | The method does not permit the clear determination of the velocities of the upstream and downstream jam fronts separately each other. This is because the method determines some average jam propagation velocity without the differentiation between the downstream and upstream jam fronts. |

# Figure captions as a list:

Figure 1: Elements of Kerner's three-phase traffic theory: (a) phase definition in measured data, (b) two-dimensional region of hypothetical steady states of synchronized flow, (c) amplitudes of critical perturbation leading to phase transitions (related to (b)) (Kerner (2004)).

Figure 2: Illustration of the FOTO and ASDA model approach (e.g., Kerner (2004, 2009b)).

Figure 3: Illustration of the M42 freeway stretch in UK (from http://www.roadtraffic-technology.com/projects/m42/m422.html).

Figure 4: Illustration Interstate I5 Northbound, Portland, USA.

Figure 5: Illustration of the A5 freeway stretch, Germany.

Figure 6: $F \rightarrow S$ transition (traffic breakdown) at a bottleneck (labelled "B") and the emergence of a wide moving jam in synchronized flow upstream of the bottleneck on UK freeway. (a) Overview of a congested pattern. (b) Average speed (left) and flow rate (right) data for middle lane of marked detectors in (a). $F \rightarrow S$ transition that occurs at detector 6429 is labelled by "$F \rightarrow S$". The upstream propagating wide moving jam that has emerged in synchronized flow is labelled by "Jam".

Figure 7: $F \rightarrow S$ transition at an USA bottleneck. (a) emergence of congested pattern, (b) averaged speed and flow data of marked detectors in (a). $F \rightarrow S$ transition at detector D20 labelled by arrow. Data from http://portal.its.pdx.edu/index.php.

Figure 8: Evolution of congested traffic pattern in 100 m distance detector measurements upstream of detector 6429: pinch effect (marked in velocity diagrams, left) on three lane freeway M42 in UK.

Figure 9: ASDA/FOTO on M42 in UK at $11^{th}$ Jan., 2008: wide moving jams propagating (dark grey) velocity of downstream front marked with thick black line.

Figure 10: ASDA/FOTO on Interstate I5-North in USA at $20^{th}$ Feb., 2009: wide moving jam propagating (dark grey) with velocity of downstream front marked with thick black line.

Figure 11: ASDA/FOTO on A5-North in Germany at $8^{th}$ Aug., 2008: wide moving jams propagating (dark grey) with velocity of downstream front marked with thick black lines.

Figure 12: ASDA/FOTO on A5-North in Germany at $16^{th}$ Oct., 2008: wide moving jam propagating (dark grey) with velocity of downstream front marked with thick black line.

Figure 13: Measured raw detector data of average speed and flow rate per minute at detectors 6389 and 6310 on $11^{th}$ January, 2008 related to Figure 9.

Figure 14: Measured raw detector data of average speed and flow rate per minute at detectors D21 and D18 on $20^{th}$ Feb., 2009 related to Figure 10.

Figure 15: Measured raw detector data of average speed and flow rate per minute at detectors D20, D11 and D6 on $8^{th}$ August, 2008 related to Figure 11.

Figure 16: Measured raw detector data of average speed and flow rate per minute at detectors D20 and D11 on 16[th] October, 2008 related to Figure 12.

Figure 17: Cross-correlation function for flow rates from detectors D11 and D20 at A5-North related to Figure 16.

Figure 18: Cross-correlation function for flow rates from detectors D11 and D20 at A5-North related to Figure 15.

Figure 19: Flow-density-planes for three situations: (a) detector 6310 related to Figure 12; (b) detector 18 related to Figure 13; (c) detector D20 related to Figure 15.

# Figures on individual pages

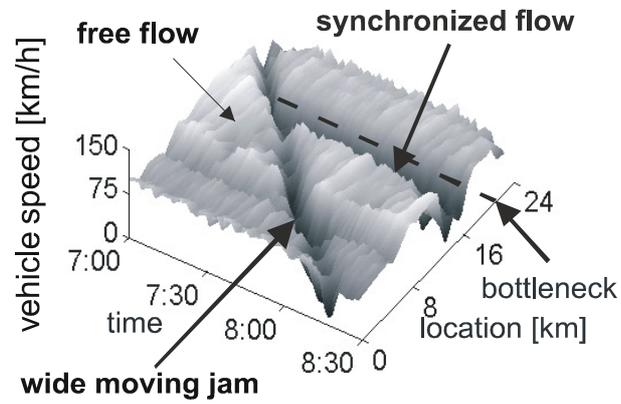
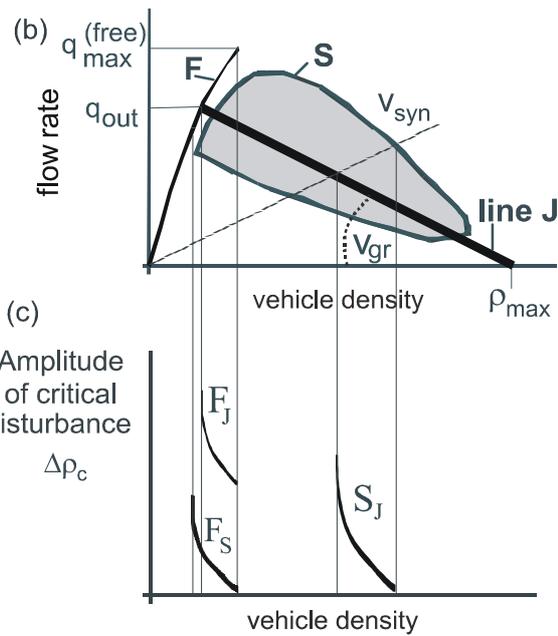

**Figure 1**. Elements of Kerner's three-phase traffic theory: (a) phase definition in measured data, (b) two-dimensional region of hypothetical steady states of synchronized flow, (c) amplitudes of critical perturbation leading to phase transitions (related to (b)) (Kerner (2004)).

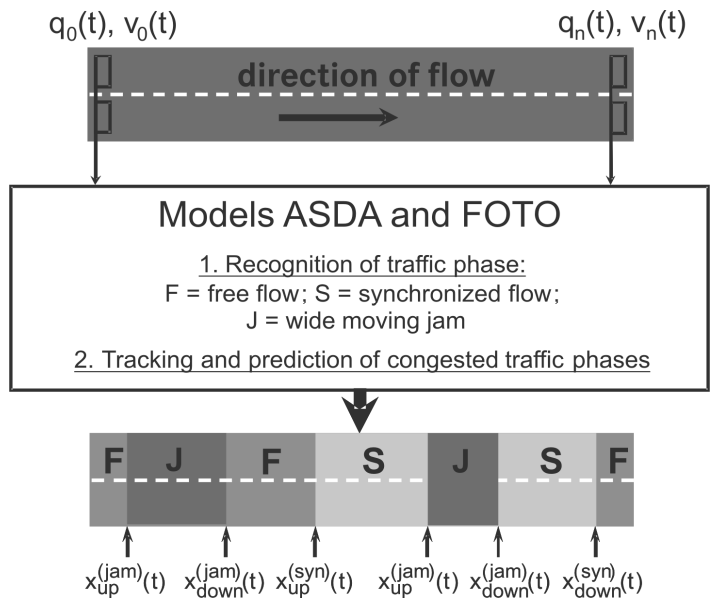

**Figure 2**. Illustration of the FOTO and ASDA model approach (e.g., Kerner (2004, 2009b)).

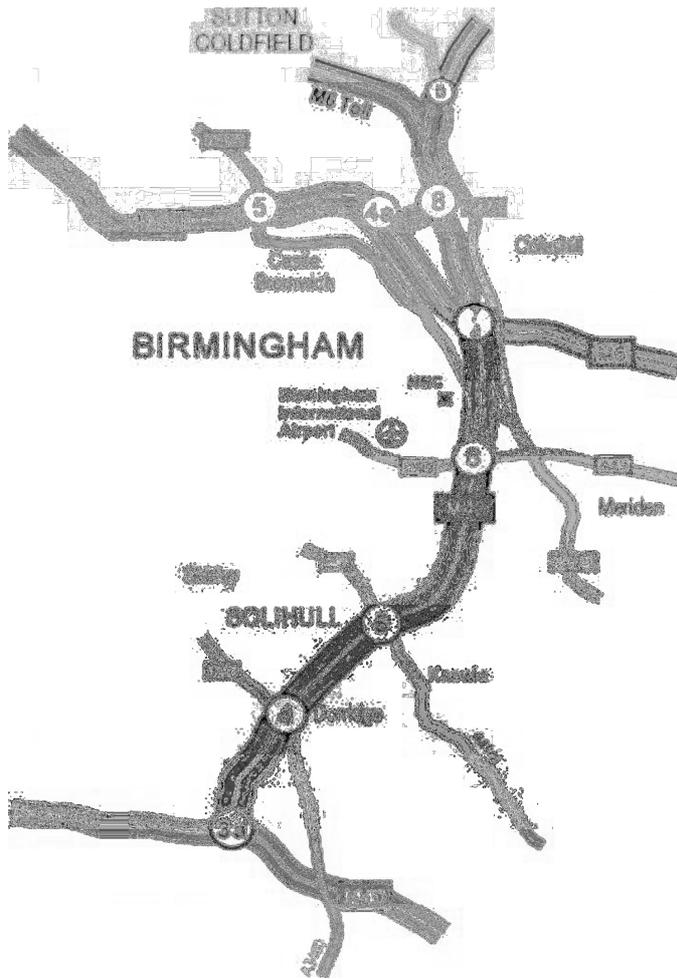

**Figure 3**. Illustration of the M42 freeway stretch in UK (from http://www.roadtraffic-technology.com/projects/m42/m422.html).

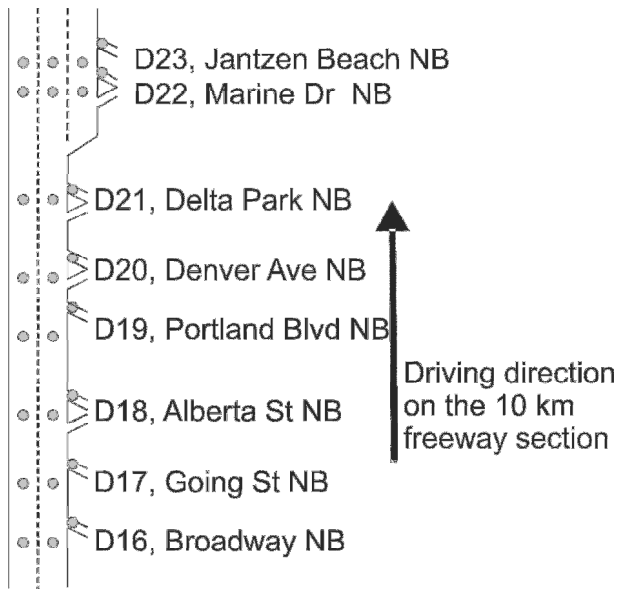

**Figure 4.** Illustration Interstate I5 Northbound, Portland, USA.

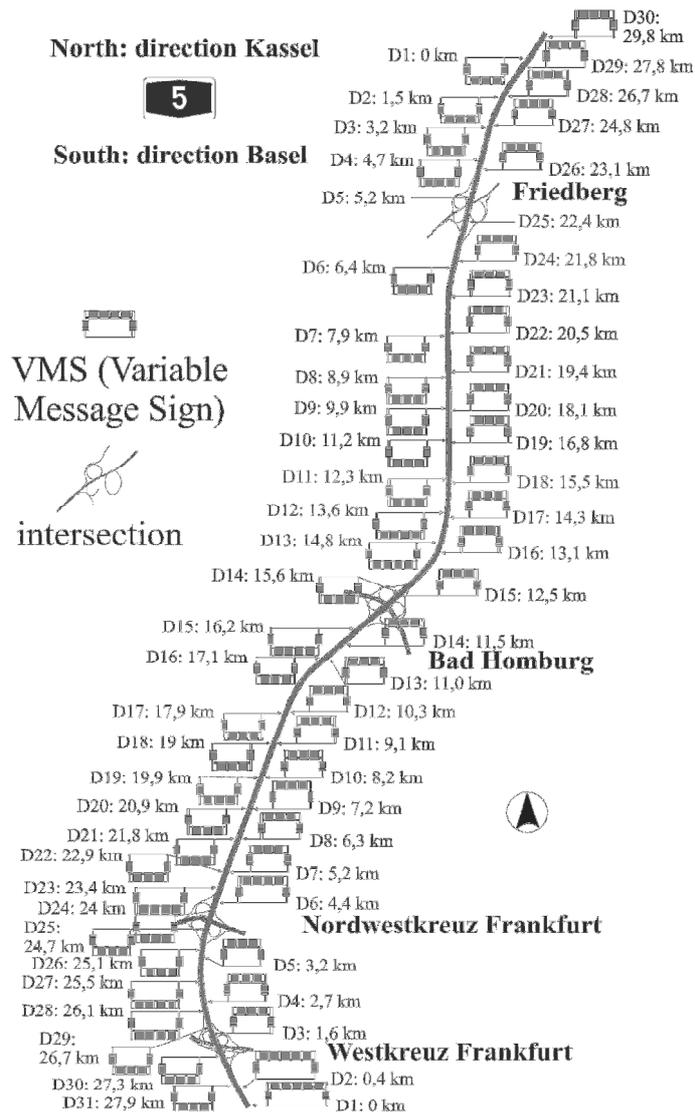

**Figure 5**. Illustration of the A5 freeway stretch, Germany.

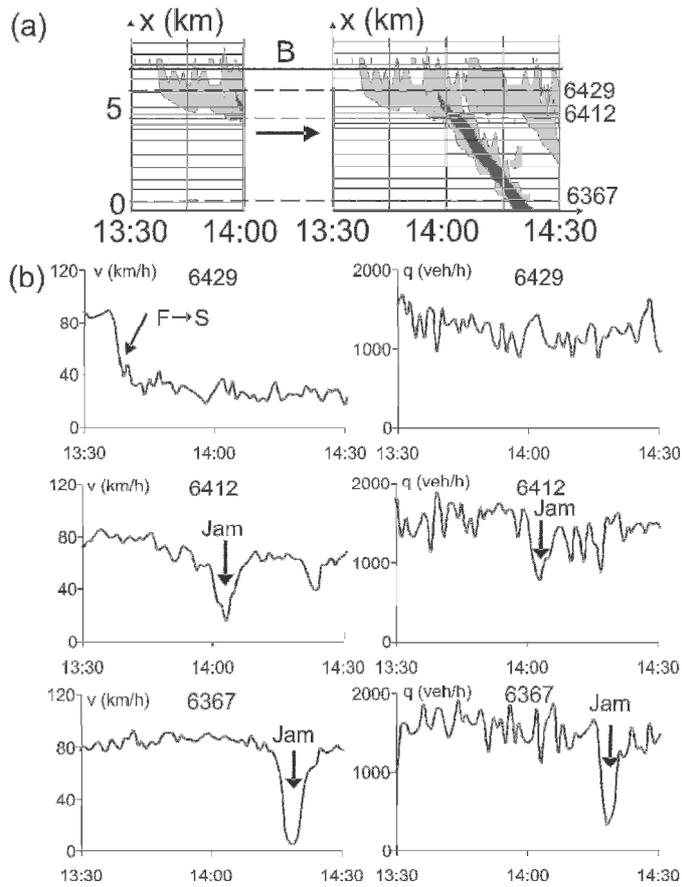

**Figure 6**. $F \to S$ transition (traffic breakdown) at a bottleneck (labelled "B") and the emergence of a wide moving jam in synchronized flow upstream of the bottleneck on UK freeway. (a) Overview of a congested pattern. (b) Average speed (left) and flow rate (right) data for middle lane of marked detectors in (a). $F \to S$ transition that occurs at detector 6429 is labelled by "$F \to S$". The upstream propagating wide moving jam that has emerged in synchronized flow is labelled by "Jam".

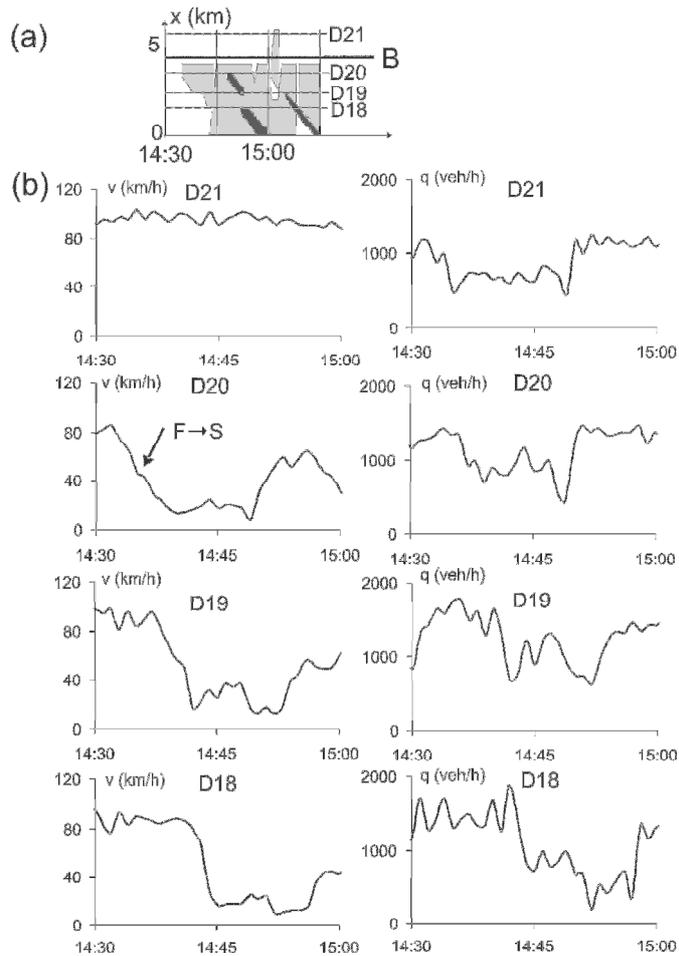

**Figure 7.** $F \rightarrow S$ transition at an USA bottleneck. (a) emergence of congested pattern, (b) averaged speed and flow data of marked detectors in (a). $F \rightarrow S$ transition at detector D20 labelled by arrow. Data from http://portal.its.pdx.edu/index.php.

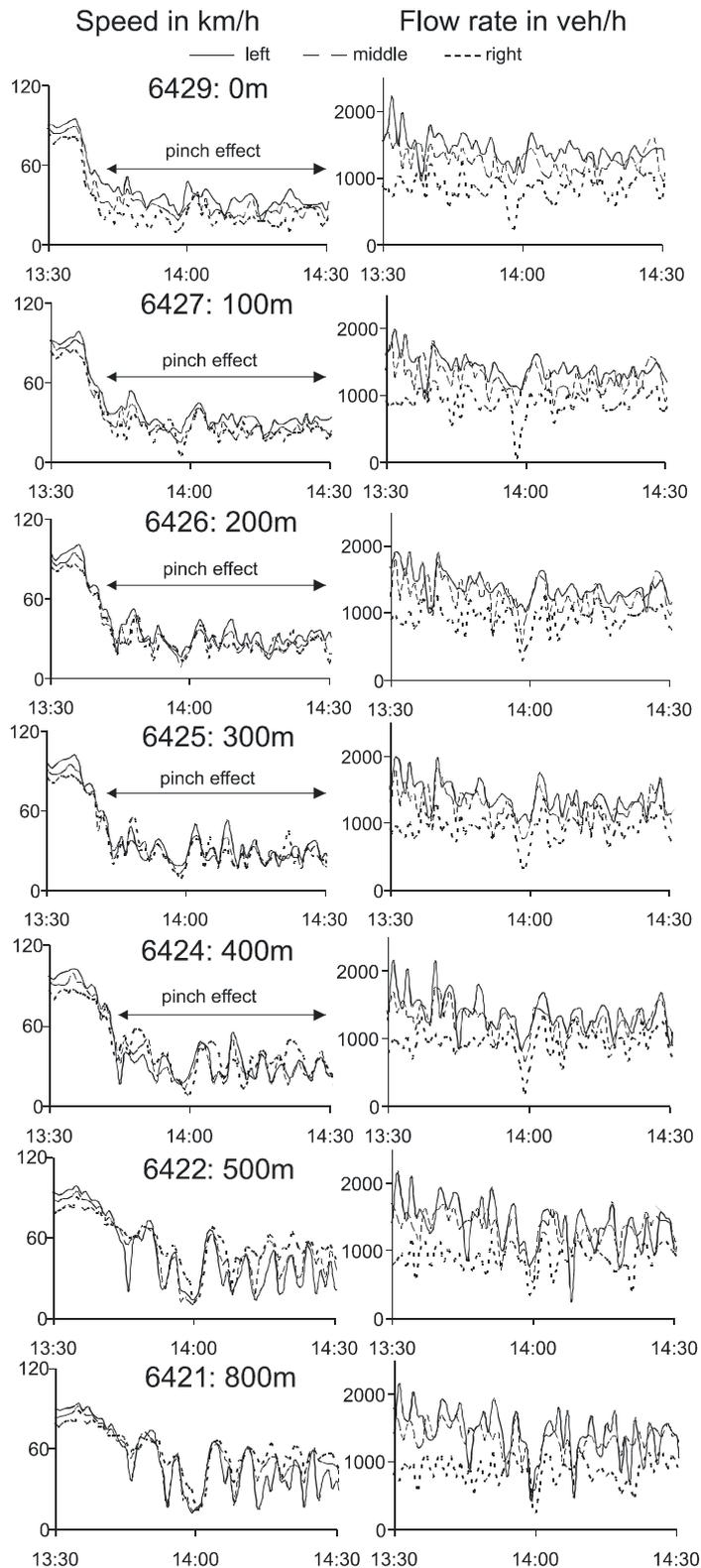

**Figure 8.** Evolution of congested traffic pattern in 100m distance detector measurements upstream of detector 6429: pinch effect (marked in velocity diagrams, left) on three lane freeway M42 in UK.

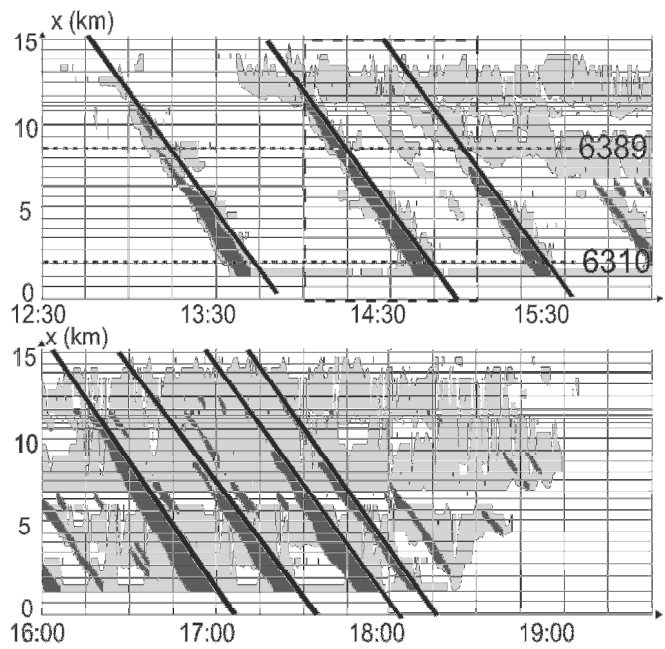

**Figure 9.** ASDA/FOTO on M42 in UK at 11[th] Jan., 2008: wide moving jams propagating (dark grey) velocity of downstream front marked with thick black line.

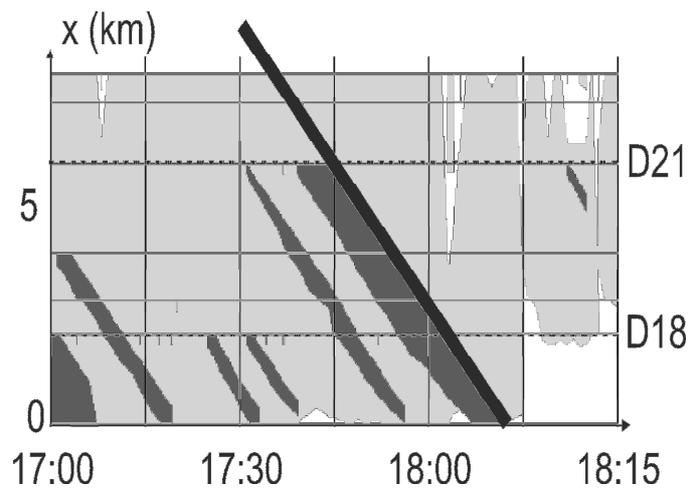

**Figure 10**. ASDA/FOTO on Interstate I5-North in USA at 20[th] Feb., 2009: wide moving jam propagating (dark grey) with velocity of downstream front marked with thick black line.

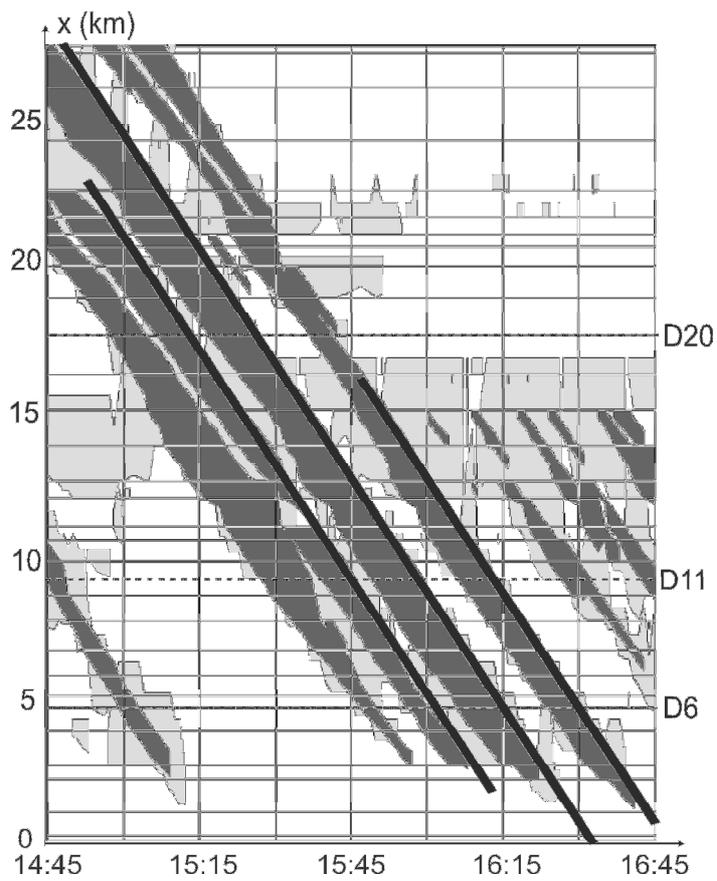

**Figure 11**. ASDA/FOTO on A5-North in Germany at 8[th] Aug., 2008: wide moving jams propagating (dark grey) with velocity of downstream front marked with thick black lines.

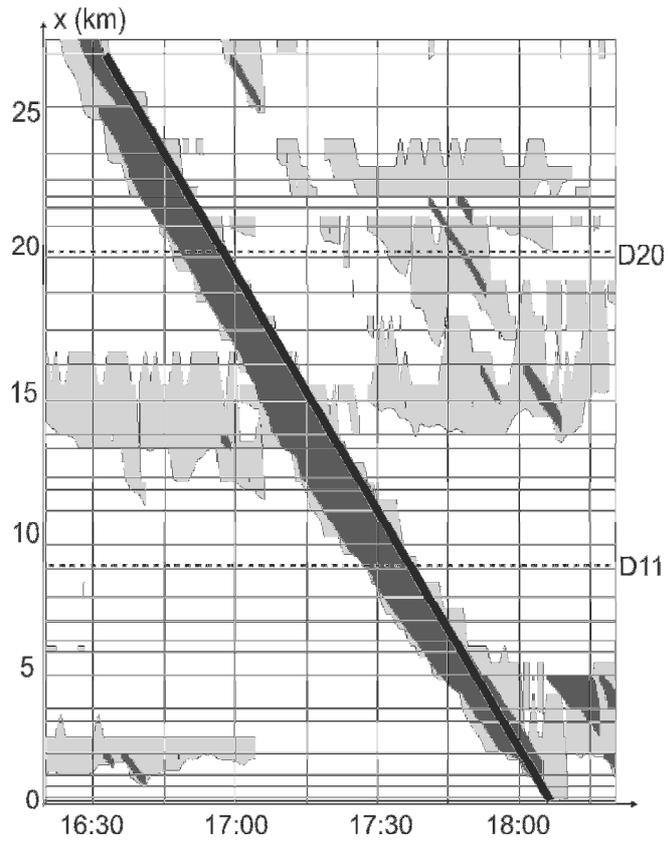

**Figure 12**. ASDA/FOTO on A5-North in Germany at 16[th] Oct., 2008: wide moving jam propagating (dark grey) with velocity of downstream front marked with thick black line.

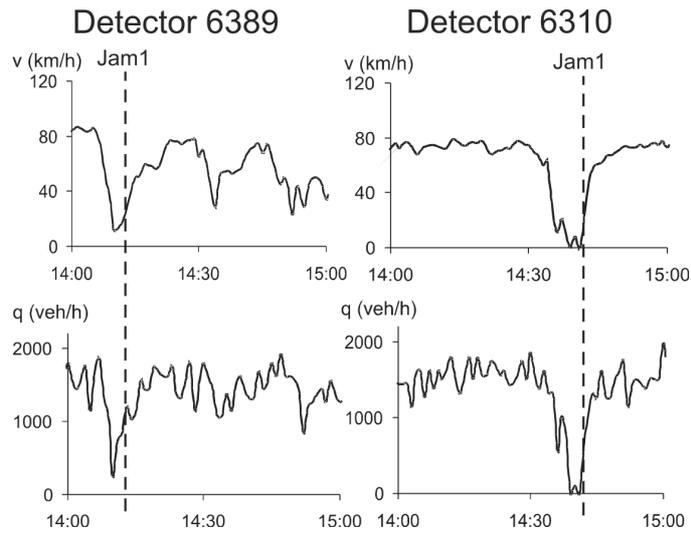

**Figure 13**. Measured raw detector data of average speed and flow rate per minute at detectors 6389 and 6310 on 11[th] January, 2008 related to Figure 9.

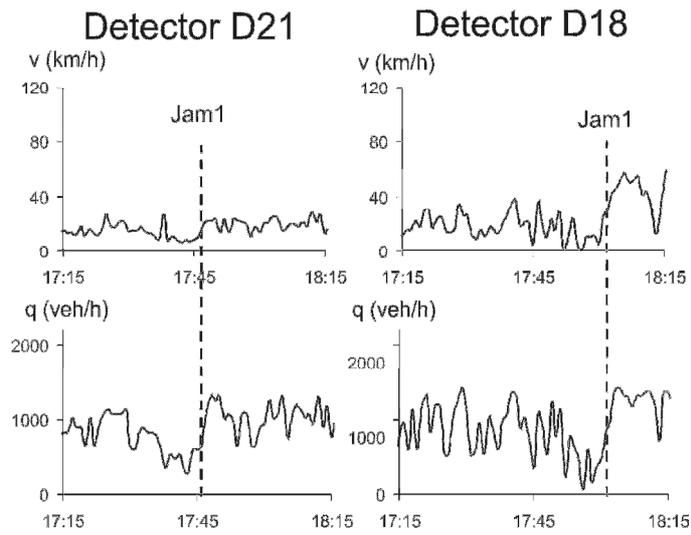

**Figure 14**. Measured raw detector data of average speed and flow rate per minute at detectors D21 and D18 on 20[th] Feb., 2009 related to Figure 10.

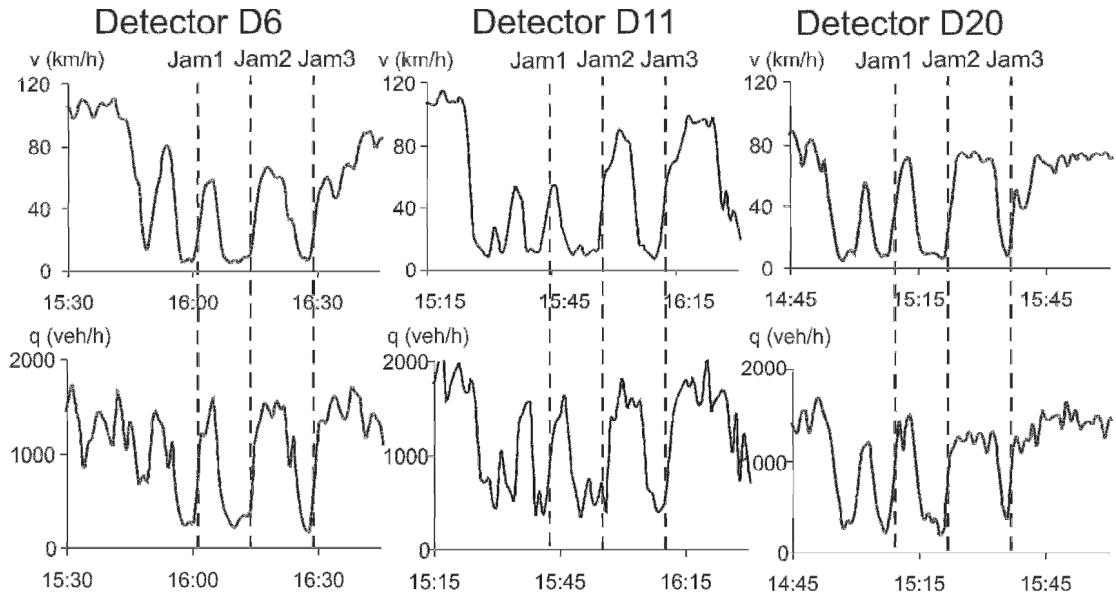

**Figure 15**. Measured raw detector data of average speed and flow rate per minute at detectors D20, D11 and D6 on 8[th] August, 2008 related to Figure 11.

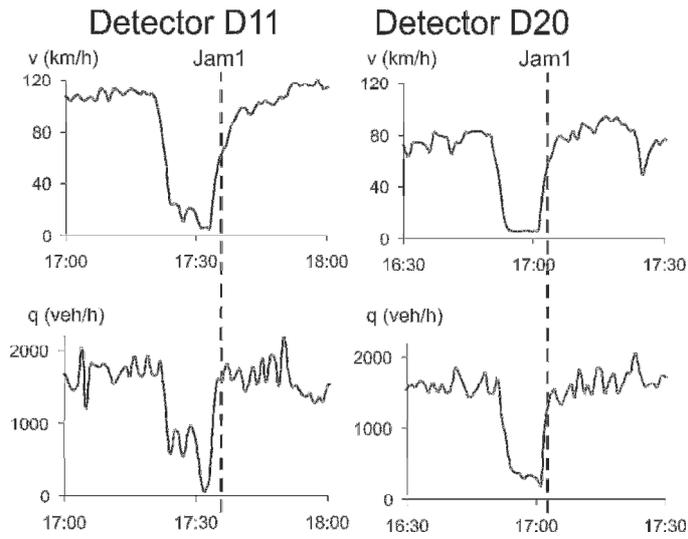

**Figure 16**. Measured raw detector data of average speed and flow rate per minute at detectors D20 and D11 on 16[th] October, 2008 related to Figure 12.

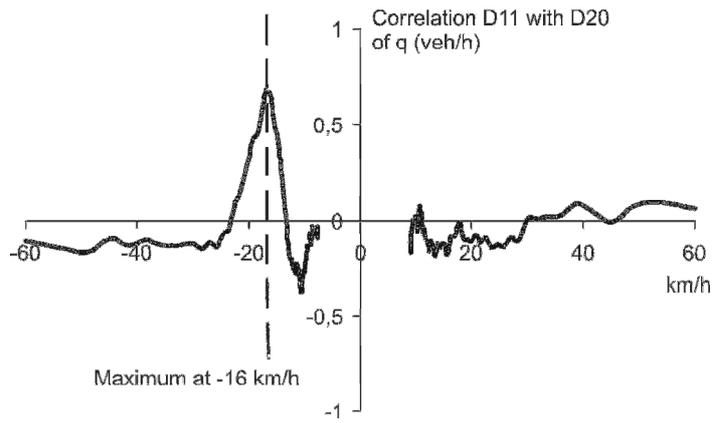

**Figure 17**. Cross-correlation function for flow rates from detectors D11 and D20 at A5-North related to Figure 16.

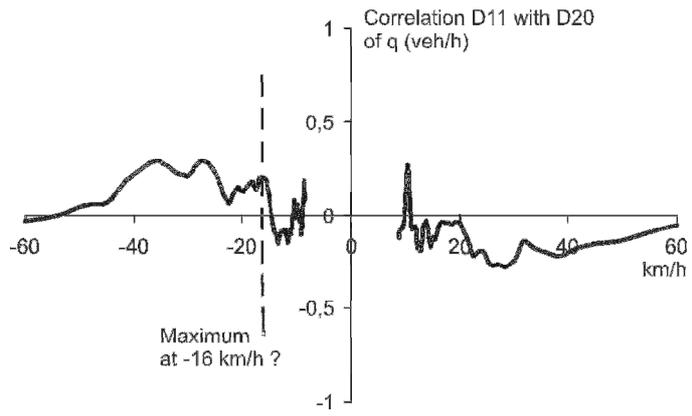

**Figure 18.** Cross-correlation function for flow rates from detectors D11 and D20 at A5-North related to Figure 15.

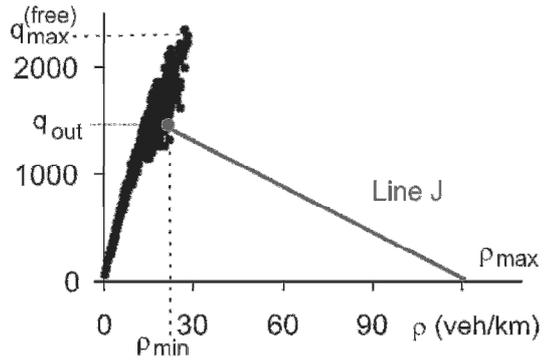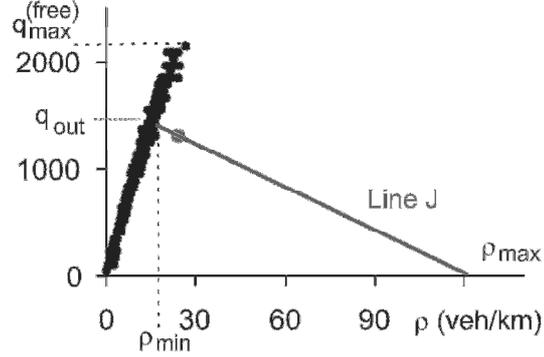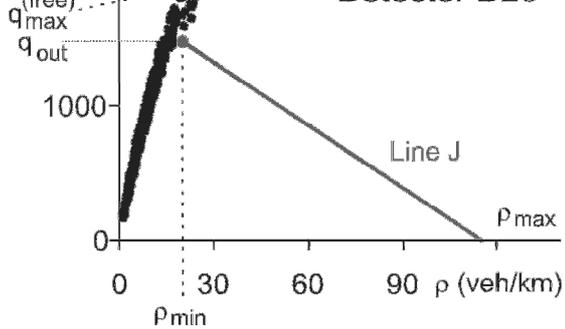

**Figure 19.** Flow-density-planes for three situations: (a) detector 6310 related to Figure 12; (b) detector 18 related to Figure 13; (c) detector D20 related to Figure 15.